\documentclass[prd, aps, superscriptaddress, preprintnumbers, twocolumn, floatfix, nofootinbib]{revtex4}
\pdfoutput=1

\usepackage{amsfonts}
\usepackage{amsmath}
\usepackage{amssymb}
\usepackage{bm}
\usepackage{dcolumn}
\usepackage{graphicx}   
\usepackage[latin1]{inputenc}
\usepackage{latexsym}
\usepackage{rotating}
\usepackage{hyperref}
\usepackage{graphicx}
\usepackage{color}

\newcommand\be{\begin{equation}}
\newcommand\ba{\begin{eqnarray}}
\newcommand\ee{\end{equation}}
\newcommand\ea{\end{eqnarray}}

\begin{document}

\title{Nonsingular Ekpyrotic Cosmology with a Nearly Scale-Invariant Spectrum of Cosmological
Perturbations and Gravitational Waves}

\author{Robert Brandenberger}
\email{rhb@physics.mcgill.ca}
\affiliation{Department of Physics, McGill University, Montr\'{e}al, QC, H3A 2T8, Canada}

\author{Ziwei Wang}
\email{ziwei.wang@mail.mcgill.ca}
\affiliation{Department of Physics, McGill University, Montr\'{e}al, QC, H3A 2T8, Canada}

\date{\today}

\begin{abstract}

We  propose a mechanism borrowed from string theory which yields a non-singular transition from a phase of Ekpyrotic contraction to the expanding phase of Standard Big Bang cosmology. The same mechanism converts the initial vacuum spectrum of cosmological fluctuations before the bounce into a scale-invariant one, and also changes the spectrum of gravitational waves into an almost scale-invariant one. The scalar and tensor tilts are predicted to be the same, in contrast to the predictions from the ``String Gas Cosmology'' scenario. The amplitude of the gravitational wave spectrum depends on the ratio of the string scale to the Planck scale and may be in reach of upcoming experiments.

\end{abstract}

\maketitle

\section{Introduction} 
\label{sec:intro}

The Inflationary Universe scenario \cite{inflation} has become the standard paradigm of early universe cosmology. It is based on the assumption that there was a period of almost exponential expansion during a time period in the very early universe. Inflation provides a solution of the horizon and flatness problems of Standard Big Bang cosmology, and provides a causal mechanism for producing cosmological perturbations and microwave background anisotropies based on the assumption that all fluctuation modes start our in their vacuum state inside the Hubble radius at early times \cite{Mukh}. The spectrum of curvature perturbations is predicted to be almost scale-invariant, with a slight red tilt. Inflation also produces \cite{Starob} an approximately scale-invariant spectrum of gravitational waves, again with a slight red tilt.

Inflation is usually obtained by working in the context of Einstein gravity and assuming that there is a new scalar field, the {\it inflaton} field $\varphi$, whose stress-energy tensor has an equation of state 
\be
w \, \equiv \, \frac{p}{\rho} \, \simeq \, -1
\ee
(where $p$ and $\rho$ are pressure and energy density, respectively) which leads to accelerated expansion. 

Recently, however, inflationary cosmology has come under some pressure. First of all, recall that in order for a scalar field $\varphi$ to serve as an inflaton, its potential energy $V(\varphi)$ has to be very flat in order that the potential energy dominates over the kinetic energy \footnote{Warm inflation \cite{Berera} provides an avenue of relaxing this constraint.}. However, general arguments from string theory lead to the {\it swampland constraint} \cite{Vafa} (see \cite{Palti} for reviews)
\be
\frac{V^{\prime}}{V} \, > \frac{c}{m_{pl}} 
\ee
for slowly rolling scalar fields whose energy density dominates the universe, where $c$ is a constant of the order $1$, $m_{pl}$ is the four space-time dimensional Planck mass, and a prime indicates the derivative with respect to the field $\varphi$. Effective field theory models which violate this constraint are said to lie in the {\it swampland} and are not consistent with string theory \footnote{These considerations also have implications for scalar field models of Dark Energy \cite{Lavinia}.}. Single scalar field models of slow-roll inflation are thus \cite{AOSV} in the swampland \footnote{Once again, models of warm inflation avoid this constraint \cite{warm2}.}.

A second constraint on inflationary cosmology comes from the recently proposed {\it Trans-Planckian Censorship} conjecture (TCC) \cite{TCC} which states that during cosmological evolution no scales whose wavelengths were smaller than the Planck length ever exit the Hubble horizon. This conjecture can be viewed as analogous to Penrose's {\it Cosmic Censorship} hypothesis \cite{Penrose} which states that timelike singularities must be hidden by horizons. If the TCC is satisfied, then trans-Planckian modes are hidden from the classical region \footnote{Fluctuation modes oscillate on sub-Hubble scales, but become squeezed states and classicalize on super-Hubble scales \cite{Kiefer} (see \cite{MFB} for reviews of the theory of cosmological perturbations).}. The TCC also shields the classical region of cosmology from non-unitarities associated with setting up quantum field theory in an expanding background \cite{Weiss}. Since during inflation the physical wavelength of fluctuation modes increases almost exponentially, while the Hubble radius remains almost constant, the TCC clearly provides severe constraints on inflationary cosmology. In \cite{TCC2} it was shown that, assuming that the post-inflationary cosmology is like in Standard Big Bang cosmology, and that the potential energy during inflation is approximately constant, the potential energy is constrained to obey the upper bound
\be
V^{1/4} \, < \, 3 \times 10^{9} {\rm{Gev}} \, ,
\ee
which leads to an upper bound on the tensor-to-scalar ratio $r$ of
\be
r \, < \, 10^{-30} \, .
\ee
Even if these constraints are enforced, an initial condition problem remains \cite{TCC2}.

In light of these constraints on inflationary cosmology it is interesting to re-consider some alternative early universe scenarios. Any viable alternative to inflation should produce an approximately scale-invariant spectrum of almost adiabatic cosmological fluctuations on scales which are being observed today. In this case, as shown in the pioneering papers \cite{SZ, Peebles}, acoustic oscillations in the angular power spectrum of the Cosmic Microwave Background (CMB) and baryon acoustic oscillations in the matter power spectrum will be generated. Two promising classes of alternative scenario (see e.g. \cite{RHBrev} for a review and comparison of these alternatives) are {\it bouncing} and {\it emergent} cosmologies. In bouncing cosmologies (see e.g. \cite{bouncerev} for reviews) it is assumed that the universe begins in a contracting phase, and new physics produces a bounce which leads to the current expanding phase of Big Bang cosmology. Fluctuations are taken to be in their vacuum state in the far past. In the {\it emergent} scenario, it is assumed that the current phase of cosmological expansion starts after a phase transition from a novel state of space-time-matter. One example is {\it String Gas Cosmology} \cite{BV} where it is assumed that the early phase is a hot gas of fundamental superstrings near the critical temperature of string theory, and thermal fluctuations with holographic scaling in this hot gas lead to an almost scale-invariant spectrum of curvature fluctuations \cite{NBV} with a slight red tilt and an almost scale-invariant spectrum of gravitational waves \cite{BNPV} with a slight blue tilt. Provided that the energy scale of the bounce or of the emergent phase is lower than the Planck scale, there are no constraints on the scenarios resulting from the TCC. Note that in bouncing and emergent scenarios the horizon problem is trivially solved - the causal horizon is infinite, and there is hence in principle no problem in explaining the near isotropy of the CMB.

Among bouncing scenarios, the {\it Ekpyrotic Scenario} \cite{Ekp} (see also \cite{cyclic} for a cyclic version) has a number of attractive features. The Ekpyrotic scenario assumes that the contracting phase has an equation of state parameter
\be
w \, \gg \, 1 \, .
\ee
This can be realized if matter is given by a scalar field $\varphi$ with negative exponential potential
\be \label{pot}
V(\varphi) \, = \, - V_0 e^{- \sqrt{2/p} \varphi / m_{pl}} \, 
\ee
with $V_0 > 0$ and $0 < p \ll 1$, and assuming that $\varphi$ begins at positive values with positive total energy density. In this case, the scale factor evolves as
\be \label{SF}
a(t) \, \sim (-t)^p \, 
\ee
(note that $t$ is negative in the contracting phase) and
\be
w \, \simeq \, \frac{4}{3p} \, .
\ee
Note that negative exponential potentials arise rather generically in string compactifications (see e.g. \cite{Baumann} for a review). The initial Ekpyrotic model was in fact based on heterotic M-theory \cite{HW} (see also \cite{Perry})  \footnote{For newer versions of the Ekpyrotic scenario see e.g. \cite{Ijjas1, Ijjas2}. Our discussion, however, will be based on the original scenario,}.

A nice feature of the Ekpyrotic scenario is that anisotropies are diluted in the phase of contraction \cite{ES}, unlike what happens in a symmetric bounce where the anisotropies blow up \cite{Peter}. A further nice feature is that the homogeneous contracting trajectory is a local attractor in initial condition space. This feature is shared by models of large field inflation \cite{LFI}, but not models of small field inflation \cite{Piran}. As in inflationary cosmology, spatial curvature is diluted. Hence, the Ekpyrotic scenario also solves the flatness problem of Big Bang cosmology \footnote{Note that some of these features are shared by the Pre-Big-Bang scenario \cite{PBB} which is based on a phase of contraction with $w = 1$.}.

The Ekpyrotic scenario faces two main challenges. The first is how to obtain a non-singular bounce from the early contracting phase to the late time expanding phase of Standard Big Bang cosmology. The second is how to obtain a roughly scale-invariant spectrum of curvature fluctuations. It can be shown that the adiabatic curvature fluctuations in a phase of Ekpyrotic contraction retain a nearly vacuum spectrum \cite{Ekpflucts} in spite of the fact that the spectrum of fluctuations of the scalar field $\varphi$ obtains a scale-invariant spectrum \cite{KOST2}. It is possible to obtain a scale-invariant spectrum of curvature fluctuations making use of en entropy field which acquires a scale-invariant spectrum \cite{newEkp}, and converting the entropy fluctuations to curvature perturbations \footnote{An almost scale-invariant spectrum can also be obtained \cite{Khoury} by making use of a rapidly changing equation of state before the phase of Ekpyrotic contraction.}. It has also been shown in \cite{DV} that non-trivial matching conditions of fluctuations on the space-like surface separating the contracting phase from the expanding one can convert the scale-invariant spectrum of $\varphi$ to that of curvature perturbations \footnote{In the case of a singular bounce between an Ekpyrotic contracting phase and an expanding phase the transfer of fluctuations was studied in \cite{Tolley} (see also \cite{AdS} for a study of the transfer in a holographic cosmology setup.)}. The spectrum of gravitational waves, on the other hand, remains close to vacuum, and hence a negligible amplitude of such waves on cosmological scales is predicted.

In this paper we suggest a way of simultaneously obtaining a cosmological bounce and obtaining a scale-invariant spectrum of both curvature fluctuations and gravitational waves \footnote{See \cite{Galileon} for attempts at obtaining a nonsingular Ekpyrotic bounce using a cubic Galileon Lagrangian.}. Our mechanism is based on the fact that (in the context of string theory), at the string scale, enhanced symmetries in the low energy effective action are expected to appear (see e.g. \cite{Watson}): a tower of string states which has string scale mass in a Minkowski space-time background becomes massless and has to be included in the low energy effective action \footnote{This is the same physics discussed under the name {\it distance conjecture} in the recent superstring literature \cite{distance}.}. In the low energy effective action, there is thus an extra term which appears at the time $t = t_B$ when the density reaches the string scale density. It is a delta function term localized on the space-like hypersurface $t = t_B$ (as will be discussed in Section 3 we are working in the uniform density gauge), and we hence call this term an {\it S brane}. In the same way that a D-brane has negative pressure (equals positive tension) in the spatial directions along the brane and vanishing pressure in the normal direction, an S-brane has vanishing energy density and negative pressure. This is discussed in detail in \cite{Kounnas} in the context of a specific string model with thermal duality in the Euclidean temporal direction. The S-brane hence yields a contribution to the effective energy-momentum tensor which violates the Null Energy Condition and hence, as shown in \cite{Kounnas} can meditate a transition from contraction to expansion. As we show here, the effect of the S-brane on the cosmological perturbations and gravitational waves converts initial vacuum fluctuations before the bounce to scale-invariant ones after the bounce.

In the following section we discuss the origin of the S-brane and how this object mediates the transition between contraction and expansion. Then, in Section 3 we study the coupling of cosmological perturbations and gravitational waves to the S-brane and show that the slope of the power spectrum of curvature perturbations changes by a factor of $k^{-2}$, $k$ being comoving momentum. Thus, the vacuum power spectrum with a small red tilt $\delta n$ produced during the Ekpyrotic phase of contraction is converted into a scale-invariant one with the same small red tilt $\delta n$. The spectrum of gravitational is enhanced by the same mechanism when passing through the bounce. Thus, unlike in the other approaches to Ekpyrotic cosmology \cite{newEkp}, we obtain a roughly scale-invariant spectrum of gravitational waves \footnote{Note that the anamophic scenario of \cite{Ijjas2} also produces an approximately scale-invariant spectrum of gravitational waves.}.

Our discussion is at the level of an effective field theory, but we have in mind a setting coming from string theory. We will work in units where the speed of light and the Planck and Boltzmann constants are set to $1$. Space-time indices are denoted by Greek symbols, spatial indices by lower case Latin ones. The cosmological scale factor is denoted by $a(t)$, and $H(t)$ is the Hubble expansion rate. $G$ is Newton's gravitational constant, related to the reduced Planck mass $m_{pl}$ via $8 \pi G = m_{pl}^{-2}$. Since spatial curvature is not important in the early universe we will set it to zero.
 
\section{S-Brane and Nonsingular Bounce}

We are working with a four-dimensional effective action $S$ of the form
\ba \label{action}
S \, &=& \, \int d^4x \sqrt{-g} \bigl[ R + \frac{1}{2} \partial_{\mu} \varphi \partial^{\mu} \varphi
- V(\varphi) \bigr] \nonumber \\
& & - \int d^4x \kappa \delta(\tau - \tau_B) \sqrt{\gamma} \, ,
\ea
where $R$ is the Ricci scalar of the four-dimensional space-time metric $g_{\mu \nu}$ with determinant $g$, $\varphi$ is the scalar field with negative exponential potential $V(\varphi)$ (\ref{pot}), $\gamma_{ij}$ is the induced metric on the hypersurface $t = t_B$ with determiant $\gamma$, and $\kappa$ is the tension of the S-brane. One of the conditions on our coordinates (this will be important when studying fluctuations in the next section) is that the constant $t$ surfaces correspond to constant density. The time $t_B$ is the time when the density reaches the critical value when the extra tower of string states becomes massless and when hence the S-brane appears.

As discussed in detail in \cite{Kounnas}, the S-brane induces a localized stress-energy tensor with energy density $\rho_B$ and pressure $p_B$ given by
\ba
\rho_B \, &=& \, 0 \, , \\
p_B \, &=& \, - \kappa  \delta(t - t_B) \, .
\ea

Integrating the Friedmann equations across the bounce time $t_B$ yields the following change of the Hubble constant:
\ba
\delta H \, & \equiv & \,  \lim_{\epsilon \rightarrow 0} H(t_B + \epsilon) - H(t_B - \epsilon) \nonumber \\
& = & 4 \pi G \kappa \, .
\ea
Hence, provided that the energy density just before the bounce obeys the constraint
\be
\rho(t_B)^{1/2} \, < \, \frac{\sqrt{3}}{2} m_{pl}^{-1} \kappa \, ,
\ee
then the S-brane will induce a cosmological bounce. We expect the S-brane to appear at the string energy scale $\eta_s$, and hence $\rho(t_B) \sim \eta_s^4$. We expect $\kappa$ to be given by 
\be
\kappa \, \sim \, N \eta_s^3 \, ,
\ee
where $N$ is an integer given by the number of string states which become massless at the enhanced symmetry point (where the brane appears). Hence, provided that the $N \eta_s$ is larger than the Planck mass, the S-brane will induce a cosmological bounce.

We will assume that small-scale interactions between the enhanced symmetry states and the degrees of freedom of the Standard Model lead to a bath of radiation after the bounce. This is the analog of the reheating process at the end of inflation. With this assumption, the bounce will induce a transition between an Ekpyrotic phase of contraction and a radiation phase of expansion after the bounce.

\section{Fluctuations Passing Through the S-Brane Bounce}

Let us now consider how metric fluctuations couple to the S-brane. In this section it is convenient to use conformal time $\tau$ in terms of which the homogeneous and isotropic background metric takes the form
\be
ds^2 \, = \, a^2(\tau) \bigl( d\tau^2 - d{\bf x}^2 \bigr) \, .
\ee
We will consider scalar metric fluctuations and gravitational waves separately \footnote{We postpone the discussion of vector modes to a later study.}. The metric for scalar fluctuations takes the form (see e.g. \cite{MFB} for reviews of the theory of cosmological perturbations)
\be
g_{\mu \nu} \, = \, a^2(\tau)
\begin{pmatrix}
1 + 2\Phi & - B_{, i} \\
- B_{, i} & (1 + 2\Psi)\delta_{ij} + E_{, ij} 
\end{pmatrix}
\, ,
\ee
where the fluctuation variables $\Phi$, $\Psi$, $B$ and $E$ are functions of space and time. The scalar field is
\be
\varphi(\tau, {\bf x}) \, = \, \varphi_0(\tau) + \delta \varphi({\bf x}, \tau) \, ,
\ee
where $\varphi_0$ is the background scalar field and $\delta \varphi$ is the field fluctuation.

Not all of the fluctuation variables are independent. We will work in comoving gauge 
$\delta \varphi = 0$ in which the scalar field energy density is constant on constant time surfaces (on large scales where the spatial gradient energy is negligible). We can impose a second gauge condition and choose it to be $E = 0$ for computational ease.

We wish to obtain the effects of the S-brane on the equation of motion for the fluctuations. To this end, we insert the above ansatz for the perturbated metric and perturbed matter into the full action (\ref{action}) and expand to second order in the fluctuation variables. Note that the terms linear in the fluctuations vanish if the background satisfies the background equations of motion. The contribution of the bulk term in the action to the second order action for scalar fluctuations is
\be
S^{(2)} \, = \, \frac{1}{2} \int d^4x \bigl[ v'^2 - v_{, i}v_{, i} + {\frac{z''}{z}} v^2 \bigr] \, ,
\ee
where $v$ is the Mukhanov-Sasaki variable \cite{MukhSas} which is given by
\be
v \, = \, z \zeta \, ,
\ee
where $\zeta$ is the curvature fluctuation in comoving gauge, and 
\be
z(\tau) \, \propto \, a(\tau)
\ee
if the equation of state of the background is time-independent. A prime in the above equations denotes a derivative with respect to conformal time. 

During the phase of Ekpyrotic contraction we have (see (\ref{SF}))
\be
\tau(t) \, = \, - \frac{1}{1 - p} (-t)^{1 - p} \, ,
\ee
and hence
\be
z(\tau) \, \sim \, \tau^{p / (1 - p)} \, ,
\ee
from which it follows that
\be
\frac{z''}{z} \, = \, \frac{p (2p -1)}{(1 - p)^2} \tau^{-2} \, ,
\ee
which implies that an initial vacuum spectrum remains almost vaccum on super-Hubble scales, acquiring only a small red tilt proportional to $p$. After the bounce in the radiation phase of expansion we have
\be
\frac{z''}{z} \, = \, 0 \, ,
\ee
as is well known.

In the gauge we are using the induced metric $\gamma_{ij}$ is
\be
\gamma_{ij} \, = \, a^2 (1 + 2 \Psi)\delta_{ij} \, ,
\ee
and hence the contribution $S^{(2)}_B$ of the brane term in the action to quadratic order in $\Psi$ is
\be
S^{(2)}_B \, = \, \int d^4x \kappa a^3 (1 + 3 \Psi + \frac{3}{2} \Psi^2) \delta(\tau - t\tau_B) \, .
\ee
Note that the sum of the terms linear in the fluctuations cancel since we are expanding about a solution of the background equations.

In the gauge we are using $\Psi$ is proportional to the Sasaki-Mukhanov variable $v$:
\be
\Psi \, = \, z^{-1} v \, ,
\ee
where $z$ is a function of the background cosmology which depends both on the geometry and on the matter (its form will be discussed below), and hence the brane contribution to the second order action can be re-written as
\be
S^{(2)}_B \, = \, \frac{1}{2} \int d^4x {\frac{a^3}{z^2}} 3 \kappa \delta(\tau - \tau_B) v^2 \, ,
\ee
and thus the full second order action for $v$ is
\be
S^{(2)} \, = \, \frac{1}{2} \int d^4x \bigl[ v'^2 - v_{, i} v_{, i} + 
\bigl( {\frac{z''}{z}} + 3 {\frac{a^3}{z^2}} \kappa \delta(\tau - \tau_B) 
\bigr) v^2 \bigr] \, ,
\ee
and the resulting equation of motion for $v$ is
\be \label{veq}
v'' \, + \, \bigl[ k^2 - {\frac{z''}{z}} - 3 {\frac{a^3}{z^2}} \kappa \delta(\tau - \tau_B) 
\bigr] v \, = \, 0 \, .
\ee
Since $z'' = 0$ after the bounce and $z''/z$ is proportional to $p$ and hence very small before the bounce, we can to first approximation neglect this term.

As shown in the Appendix, the $\delta$ function contribution to the mass in the equation of motion (\ref{veq}) leads to an enhancement of the mode functions by a factor 
\be
\beta_k \, = \, \frac{m}{k} \, , 
\ee
where $m$ is the coefficient of the delta function. From (\ref{veq}) it follows that
\be \label{scalarresult}
\beta_k \, = \, 3 \kappa a \bigl( \frac{a}{z} \bigr)^2(\tau_B) k^{-1} \, .
\ee
This factor leads to a conversion of the vacuum power spectrum of $v$ before the bounce to a scale-invariant one after the bounce. This is the main result of our analysis. If the spectrum before the bounce is a vacuum spectrum modulated by a slight red tilt (as it is in the case of Ekpyrotic contraction), the power spectrum after the bounce will be a scale-invariant one modulated by the same red tilt.

Let us now study the coupling of gravitational waves to the S-brane. If we consider a gravitational wave travelling in the $x$ direction, then the induced metric $\gamma_{ij}$ is
\be
\gamma_{ij} \, = \, a^2(\tau)
\begin{pmatrix}
1  & 0 \\
0 & 1 + h \epsilon_{ab}
\end{pmatrix}
\, ,
\ee
where $\epsilon_{ab}$ is the polarization tensor of gravitational waves in the $y/z$ plane. Hence, to leading order in the amplitude $h^2$
\be
\sqrt{\gamma} \, = \, a^3 \bigl(1 - \frac{1}{2} h^2 \bigr) \, .
\ee

The canonical variable for gravitational waves is
\be
u \, \equiv \, a h m_{pl} \, ,
\ee
where the factor of $m_{pl}$ is important to have the right dimensions. The bulk action for $u$ is
\be
S^{(2)} \, = \, \frac{1}{2} \int d^4x \bigl[ u'^2 - u_{, i} u_{, i} + {\frac{a''}{a}} u^2 \bigr] \, ,
\ee
and the brane contribution is
\be
 S^{(2)}_B \, = \,   \frac{1}{2} \int d^4x \kappa a \delta(\tau - \tau_B) u^2 m_{pl}^{-2}\, ,
\ee
Hence, the equation of motion for $u_k$ becomes
\be \label{ueq}
u'' \, + \, \bigl[ k^2 - {\frac{a''}{a}} -  \kappa a m_{pl}^{-2} \delta(\tau - \tau_B) \bigr] u \, = \, 0 \, .
\ee
This equation is analogous to the equation for scalar modes, except that the coefficient of the delta function term is different. In analogy with what was shown above for the scalar modes, we find that the tensor modes are enhanced by the factor
\be \label{tensorresult}
\beta_k \, = \, \kappa a m_{pl}^{-2} k^{-1} \, .
\ee

Our results (\ref{scalarresult}) and (\ref{tensorresult}) show that the passage through the S-brane leads to the conversion of an initial vacuum spectrum for the scalars and tensors to a scale-invariant one for both scalars and tensors. This is very different from what is obtained in previous realizations of Ekpyrotic cosmology where the tensors retain their vacuum form.

The tensor to scalar ratio $r$ can be read off of the amplitude ratio of the Bogoliubov coefficients $\beta_k$. To evaluate this ratio, recall \cite{MFB} that
\be
z(t) \, = \, a(t) \frac{{\dot{\varphi_0}}}{H} \, ,
\ee
where $\varphi_0(t)$ is the background scalar field. For the Ekpyrotic potential of (\ref{pot}), this background is given by
\be
\varphi_0(t) \, = \, \sqrt{2p} m_{pl} {\rm{log}}(- \frac{\sqrt{V_0}}{m_{pl} \sqrt{p(1 - p)}} t) \, ,
\ee
and hence
\be \label{zequation}
\frac{{\dot{\varphi_0}}}{H} \, = \sqrt{2/p} m_{pl} \, .
\ee
Hence, comparing (\ref{tensorresult}) and (\ref{scalarresult}) and making use of (\ref{zequation}) we find that the tensor to scalar ratio $r$ becomes
\be \label{ratio}
r \, = 16 \pi \bigl( \frac{72}{p^3} \bigr)  \, \sim \, \frac{10^3}{p^3} \, .
\ee
The factor of $16 \pi$ stems from the different normalizations of the scalar and tensor power spectrum, the other term is the square of the ratio of the Bogoliubov coefficients (\ref{tensorresult}) to (\ref{scalarresult}). 

We find that the amplitude of the tensor spectrum is larger than the amplitude of the scalar spectrum, a result which is obviously inconsistent with observations. This is related to the fact that the analog of the inflationary slow-roll parameter $\epsilon$ is proportional to $1/p \gg 1$ in the case of Ekpyrotic contraction. Hence, in our model we need to invoke a separate mechanism to boost the scalar modes, as considered in previous work on the Ekpyrotic scenario \cite{newEkp, DV}. This will be explored in an upcoming paper.

Let us now consider the corrections to the spectrum due to the $z''/z$ term in the mode equation. Since
\be
{\frac{z''}{z}} \, = \, - \frac{p (1 - 2p)}{(1 - p)^2} \frac{1}{\tau^2} \, \equiv \, - \alpha(p) \frac{1}{\tau^2} \, ,
\ee
the dominant mode in for the $v$ and $u$ modes on super-Hubble wavelengths scales as
\be
v(\tau) \, \sim \, \tau^{\alpha} 
\ee
and similarly for $u$. Hence we find for the Fourier modes of $v$ and $u$ on super-Hubble lengths
\be \label{calc1}
v_k(\tau) \, \simeq \, v_{k, 0} \bigl( \frac{\tau}{\tau_H(k)} \bigr)^{\alpha} \, ,
\ee
where $\tau_H(k)$ is the time of Hubble radius crossing of the k'th mode and is given by
\be \label{calc2}
\tau_H(k) \, = \, \frac{p}{(1 - p)} k^{-1} \, .
\ee
If we use vacuum initial conditions then
\be \label{calc3}
v_{k, 0} \, = \, \frac{1}{\sqrt{2k}} \, .
\ee

The dimensionless power spectrum is given by
\be \label{calc4}
P_v(k) \, = \, k^3 |v_k|^2 \, .
\ee
According to the definition of the scalar tilt $n_s$ we have
\be \label{calc5}
P(k) \, \sim \, k^{n_s - 1} \, .
\ee
Combining (\ref{calc1}, \ref{calc2}, \ref{calc3}, \ref{calc4}) and (\ref{calc5}) we find that the tilt of the scalar modes before the bounce is given by
\be
n_s \, = \, 3 + 2 \alpha  \, \simeq \, 3 + 2p \, .
\ee
Since the passage through the S-brane changes the spectral tilt by -2, the tilt after the bounce is
\be
n_s  - 1 \, = \, 2 \alpha \,  \simeq \, 2p \, .
\ee
Thus, we predict a blue tilt of the scalar spectrum with the deviation of the tilt from that of scale-invariance whose magnitude is $2p$. The tensor spectrum has the same tilt. Note that the tensor index $n_t$ is defined via
\be
P(k) \, \sim \, k^{n_t} \, .
\ee
Thus, the spectrum aquires a slight blue tilt.

Our new Ekpyrotic scenario hence predicts roughly scale invariant scalar and tensor spectra with tilts which obey the consistency relation
\be \label{cons1}
n_t \, = \, n_s - 1 \, .
\ee

The amplitude of the spectrum of gravitational waves after the bounce can be found from (\ref{tensorresult}). The power spectrum of the canonical variable $u$ is
\be
P_u(k, \tau) \, \simeq \, a^2 \kappa^2 m_{pl}^{-4} \bigl( k \tau_B)^{-p/2} \, ,
\ee
and hence the power spectrum of gravitational waves becomes 
\be
P_h(k, \tau) \, \simeq \,  \kappa^2 m_{pl}^{-6} \bigl( k \tau_B)^{-p/2} \, ,
\ee
Recall from Section 2 that the tension $\kappa$ is expected to be given by the string scale $\eta_s$, and thus the amplitude ${\cal{A}}$ of the power spectrum (dropping the last factor above which represents the small red tilt) will be
\be \label{amplitude}
{\cal{A}} \, \simeq \,  N^2 \bigl( \frac{\eta_s}{m_{pl}} \bigr)^6 \, .
\ee
If the string scale is the one preferred by particle physics considerations in the early textbook \cite{GSW} on string theory, namely $\eta_s \sim 10^{17} {\rm GeV}$, then the amplitude is
\be
{\cal{A}} \, \sim \, 10^{-12} \, ,
\ee
which, using the observed amplitude of the scalar spectrum, corresponds to
\be
r \, \sim \, 10^{-3} \, .
\ee
Note, however, that this valuie depends sensitively on the ratio between the string scale and the Planck scale.

\section{Conclusions and Discussion} \label{conclusion}

It is generally expected that at string energy density scales a new tower of string states becomes massless. If this is the case, this tower of states has to be included in the low energy effective action for cosmological evolution. It will appear as a term in the effective action localized at a particular time, i.e. as an S-brane. This object has zero energy density and negative pressure, and can hence induce the transition from contraction to expansion.

We have used this S-brane construction to provide a new realization of the Ekpyrotic scenario. We have shown that the coupling of the S-brane to cosmological fluctuations and in particular to gravitational waves leads to a change in the spectral index of super-Hubble cosmological perturbations of $\delta n_s = -2$. This converts a vacuum spectrum into a scale-invariant one. Our minimal construction leads to a larger amplitude for the tensor modes than the scalar modes, and hence has to be supplemented by a separate source of cosmological perturbations. A natural way to obtain this will be presented in a followup paper.

The amplitude of the power spectrum of gravitational waves depends on the ratio of the string scale $\eta_s$ and Planck scale $m_{pl}$ (see (\ref{amplitude})). The tilt of the gravitational wave spectrum is predicted to be the same as the tilt of the scalar spectrum. Both tilts are blue.

It is interesting to compare these predictions with those obtained in {\it String Gas Cosmology} \cite{BV}, an emergent universe scenario motivated by string theory. String Gas Cosmology (SGC) also yields a roughly scale-invariant spectrum of scalar and tensor modes, and the amplitude of the power spectrum of cosmological perturbations is given by a combination of the string scale and the Planck scale, but in this case the fourth power and not the sixth power as here (see e.g. \cite{Patil} for detailed discussions). SGC yields two consistency relations between the four basic cosmological observables. In particular, the tensor tilt is $n_t = 1 - n_s$. The predicted tensor tilt is blue, as it is predicted to be in the scenario studied here. The scalar tilt, however, is red.

In this paper we have assumed that the state after the bounce is that of radiation, in a similar way that the state after reheating in inflation is that of radiation. The explicit model of ultraviolet physics that yields the production of radiation across the bounce remains to be explored. In future work we plan to study the amplitude of non-Gaussianities generated in our scenario.

\section*{Acknowledgement}

\noindent The research at McGill is supported in part by funds from NSERC and from the Canada Research Chair program. ZW also acknowledges partial support from a McGill Space Institute Graduate Fellowship and from a Templeton Foundation Grant.

\section*{Appendix}

Let us consider the equation of motion
\be \label{basic}
{X''}_k(\tau) + \bigl[ k^2 + m \delta(\tau - \tau_B) \bigr] X_k(\tau) \, = \, 0 \, .
\ee
The solutions are plane waves for $\tau < \tau_B$ and for $\tau > \tau_B$. If we denote the positive frequency solutions as $f_k$ and the negative frequency ones as $f_k^{*}$, where the $*$ indicates complex conjugation, then the solution which is pure positive frequency before $\tau_B$ can be written for $\tau > \tau_B$ as
\be
X_k \, = \, \alpha_k f_k + \beta_k f_k^{*} \, ,
\ee
where $\alpha_k$ and $\beta_k$ are the Bogoliubov mode matching coefficients which obey the relationship
\be
|\alpha_k|^2 - |\beta_k|^2 \, = \, 1 \, .
\ee

By integrating the equation (\ref{basic}) over time $\tau$ against a test function (a smooth function which decays exponentially at $\tau \rightarrow \pm \infty$) $f(\tau)$ it can be easily shown that
\be
\beta_k \, = \, \frac{m}{k} \, .
\ee
This result implies that the power spectrum of $X_k$ is boosted by a factor of $(m/k)^2$. This turns a vacuum spectrum into a scale-invariant one.


\begin{thebibliography}{99}

\bibitem{inflation}
A.~H.~Guth,
 ``The Inflationary Universe: A Possible Solution to the Horizon and Flatness Problems,''
 Phys.\ Rev.\ D {\bf 23}, 347 (1981)
 [Adv.\ Ser.\ Astrophys.\ Cosmol.\  {\bf 3}, 139 (1987)].
 doi:10.1103/PhysRevD.23.347;\\
 R.~Brout, F.~Englert and E.~Gunzig,
 ``The Creation Of The Universe As A Quantum Phenomenon,''
 Annals Phys.\  {\bf 115}, 78 (1978);\\
 A.~A.~Starobinsky,
 ``A New Type Of Isotropic Cosmological Models Without Singularity,''
 Phys.\ Lett.\ B {\bf 91}, 99 (1980);\\
 K.~Sato,
 ``First Order Phase Transition Of A Vacuum And Expansion Of The Universe,''
 Mon.\ Not.\ Roy.\ Astron.\ Soc.\  {\bf 195}, 467 (1981).

\bibitem{Mukh}
V. Mukhanov and G. Chibisov,
 ``Quantum Fluctuation And Nonsingular Universe. (In Russian),''
 JETP Lett.\  {\bf 33}, 532 (1981) [Pisma Zh.\ Eksp.\ Teor.\ Fiz.\  {\bf 33}, 549 (1981)].
 
\bibitem{Starob}
A.~A.~Starobinsky,
``Spectrum of relict gravitational radiation and the early state of the universe,''
  JETP Lett.\  {\bf 30}, 682 (1979)
  [Pisma Zh.\ Eksp.\ Teor.\ Fiz.\  {\bf 30}, 719 (1979)].
 
\bibitem{Berera}
A.~Berera,
  ``Warm inflation,''
  Phys.\ Rev.\ Lett.\  {\bf 75}, 3218 (1995)
  doi:10.1103/PhysRevLett.75.3218
  [astro-ph/9509049].
 
\bibitem{Vafa}
H.~Ooguri and C.~Vafa,
 ``On the Geometry of the String Landscape and the Swampland,''
 Nucl.\ Phys.\ B {\bf 766}, 21 (2007)
 doi:10.1016/j.nuclphysb.2006.10.033
 [hep-th/0605264];\\
 G.~Obied, H.~Ooguri, L.~Spodyneiko and C.~Vafa,
 ``De Sitter Space and the Swampland,''
 arXiv:1806.08362 [hep-th].

\bibitem{Palti}
T.~D.~Brennan, F.~Carta and C.~Vafa,
 ``The String Landscape, the Swampland, and the Missing Corner,''
 PoS TASI {\bf 2017}, 015 (2017)
 doi:10.22323/1.305.0015
 [arXiv:1711.00864 [hep-th]];\\
 E.~Palti,
 ``The Swampland: Introduction and Review,''
 arXiv:1903.06239 [hep-th].
 
 \bibitem{AOSV}
 P.~Agrawal, G.~Obied, P.~J.~Steinhardt and C.~Vafa,
  ``On the Cosmological Implications of the String Swampland,''
  Phys.\ Lett.\ B {\bf 784}, 271 (2018)
  doi:10.1016/j.physletb.2018.07.040
  [arXiv:1806.09718 [hep-th]].
  
\bibitem{Lavinia}
 L.~Heisenberg, M.~Bartelmann, R.~Brandenberger and A.~Refregier,
  ``Dark Energy in the Swampland,''
  Phys.\ Rev.\ D {\bf 98}, no. 12, 123502 (2018)
  doi:10.1103/PhysRevD.98.123502
  [arXiv:1808.02877 [astro-ph.CO]].
  
\bibitem{warm2}
S.~Das,
  ``Warm Inflation in the light of Swampland Criteria,''
  Phys.\ Rev.\ D {\bf 99}, no. 6, 063514 (2019)
  doi:10.1103/PhysRevD.99.063514
  [arXiv:1810.05038 [hep-th]];\\
M.~Motaharfar, V.~Kamali and R.~O.~Ramos,
  ``Warm inflation as a way out of the swampland,''
  Phys.\ Rev.\ D {\bf 99}, no. 6, 063513 (2019)
  doi:10.1103/PhysRevD.99.063513
  [arXiv:1810.02816 [astro-ph.CO]].
  
\bibitem{TCC}
A.~Bedroya and C.~Vafa,
  ``Trans-Planckian Censorship and the Swampland,''
  arXiv:1909.11063 [hep-th].
  
\bibitem{Penrose}
R.~Penrose,
  ``Gravitational collapse: The role of general relativity,''
  Riv.\ Nuovo Cim.\  {\bf 1}, 252 (1969)
  [Gen.\ Rel.\ Grav.\  {\bf 34}, 1141 (2002)].

 \bibitem{Kiefer}
   C.~Kiefer, D.~Polarski and A.~A.~Starobinsky,
  ``Quantum to classical transition for fluctuations in the early universe,''
  Int.\ J.\ Mod.\ Phys.\ D {\bf 7}, 455 (1998)
  doi:10.1142/S0218271898000292
  [gr-qc/9802003].
  
 \bibitem{MFB}
 V.F. Mukhanov, H.A. Feldman and R.H. Brandenberger,
 ``Theory of Cosmological Perturbations''
 Physics Reports \textbf{215}, 203 (1992);\\
 R.~H.~Brandenberger,
 ``Lectures on the theory of cosmological perturbations,''
 Lect.\ Notes Phys.\  {\bf 646}, 127 (2004)
 doi:10.1007/978-3-540-40918-25
 [hep-th/0306071].
  
\bibitem{Weiss}
N.~Weiss,
  ``Constraints on Hamiltonian Lattice Formulations of Field Theories in an Expanding Universe,''
  Phys.\ Rev.\ D {\bf 32}, 3228 (1985).
  doi:10.1103/PhysRevD.32.3228
  
\bibitem{TCC2}
 A.~Bedroya, R.~Brandenberger, M.~Loverde and C.~Vafa,
  ``Trans-Planckian Censorship and Inflationary Cosmology,''
  arXiv:1909.11106 [hep-th].
  
\bibitem{SZ}
R.~A.~Sunyaev and Y.~B.~Zeldovich, ``Small scale
fluctuations of relic radiation,'' Astrophys.\ Space Sci.\  {\bf 7}, 3
(1970). 

\bibitem{Peebles}
P.~J.~E.~Peebles and J.~T.~Yu, 
``Primeval adiabatic perturbation in an expanding universe,'' 
Astrophys.\ J.\  {\bf 162}, 815 (1970). doi:10.1086/150713 

\bibitem{RHBrev}
R.~H.~Brandenberger,
 ``Alternatives to the inflationary paradigm of structure formation,''
 Int.\ J.\ Mod.\ Phys.\ Conf.\ Ser.\  {\bf 01}, 67 (2011)
 doi:10.1142/S2010194511000109
 [arXiv:0902.4731 [hep-th]].

\bibitem{bouncerev}
M.~Novello and S.~E.~P.~Bergliaffa, 
``Bouncing Cosmologies,'' 
Phys.\ Rept.\  {\bf 463}, 127 (2008)
doi:10.1016/j.physrep.2008.04.006 
[arXiv:0802.1634 [astro-ph]];\\
R.~Brandenberger and P.~Peter,
  ``Bouncing Cosmologies: Progress and Problems,''
  Found.\ Phys.\  {\bf 47}, no. 6, 797 (2017)
  doi:10.1007/s10701-016-0057-0
  [arXiv:1603.05834 [hep-th]].
  
\bibitem{BV}
 R.~H.~Brandenberger and C.~Vafa,
 ``Superstrings In The Early Universe,'' Nucl.\ Phys.\ B {\bf 316}, 391 (1989).

\bibitem{NBV}
 A.~Nayeri, R.~H.~Brandenberger and C.~Vafa,
  ``Producing a scale-invariant spectrum of perturbations in a Hagedorn phase of string cosmology,''
  Phys.\ Rev.\ Lett.\  {\bf 97}, 021302 (2006)
  doi:10.1103/PhysRevLett.97.021302
  [hep-th/0511140].
  
\bibitem{BNPV}
R.~H.~Brandenberger, A.~Nayeri, S.~P.~Patil and C.~Vafa,
  ``Tensor Modes from a Primordial Hagedorn Phase of String Cosmology,''
  Phys.\ Rev.\ Lett.\  {\bf 98}, 231302 (2007)
  doi:10.1103/PhysRevLett.98.231302
  [hep-th/0604126].

\bibitem{Ekp}
J.~Khoury, B.~A.~Ovrut, P.~J.~Steinhardt and N.~Turok,
 ``The Ekpyrotic universe: Colliding branes and the origin of the hot big
 bang,''
 Phys.\ Rev.\ D {\bf 64}, 123522 (2001) [hep-th/0103239];\\
J.~Khoury, B.~A.~Ovrut, N.~Seiberg, P.~J.~Steinhardt and N.~Turok,
  ``From big crunch to big bang,''
  Phys.\ Rev.\ D {\bf 65}, 086007 (2002)
  doi:10.1103/PhysRevD.65.086007
  [hep-th/0108187].
  
\bibitem{cyclic}
J.~Khoury, P.~J.~Steinhardt and N.~Turok,
  ``Designing cyclic universe models,''
  Phys.\ Rev.\ Lett.\  {\bf 92}, 031302 (2004)
  doi:10.1103/PhysRevLett.92.031302
  [hep-th/0307132];\\
  P.~J.~Steinhardt and N.~Turok,
  ``Cosmic evolution in a cyclic universe,''
  Phys.\ Rev.\ D {\bf 65}, 126003 (2002)
  doi:10.1103/PhysRevD.65.126003
  [hep-th/0111098];\\
  P.~J.~Steinhardt  and N.~Turok,
  ``A Cyclic model of the universe,''
  Science {\bf 296}, 1436 (2002)
  doi:10.1126/science.1070462
  [hep-th/0111030].
  
\bibitem{Baumann}
D.~Baumann and L.~McAllister,
  ``Inflation and String Theory,''
  doi:10.1017/CBO9781316105733
  arXiv:1404.2601 [hep-th].

\bibitem{HW} 
P.~Horava and E.~Witten, 
``Eleven-dimensional supergravity on a manifold with boundary,'' 
Nucl.\ Phys.\ B {\bf 475}, 94 (1996)
doi:10.1016/0550-3213(96)00308-2 
[hep-th/9603142];\\ 
  P.~Horava and E.~Witten, 
  ``Heterotic and type I string dynamics from eleven-dimensions,'' 
  Nucl.\ Phys.\ B {\bf 460}, 506 (1996)
  doi:10.1016/0550-3213(95)00621-4 
  [hep-th/9510209]. 
  
\bibitem{Perry}
 N.~Turok, M.~Perry and P.~J.~Steinhardt,
  ``M theory model of a big crunch / big bang transition,''
  Phys.\ Rev.\ D {\bf 70}, 106004 (2004)
  Erratum: [Phys.\ Rev.\ D {\bf 71}, 029901 (2005)]
  doi:10.1103/PhysRevD.71.029901, 10.1103/PhysRevD.70.106004
  [hep-th/0408083].
  
\bibitem{Ijjas1}
A.~Ijjas and P.~J.~Steinhardt,
  ``A new kind of cyclic universe,''
  Phys.\ Lett.\ B {\bf 795}, 666 (2019)
  doi:10.1016/j.physletb.2019.06.056
  [arXiv:1904.08022 [gr-qc]].
  
 \bibitem{Ijjas2} 
  A.~Ijjas and P.~J.~Steinhardt,
  ``The anamorphic universe,''
  JCAP {\bf 1510}, 001 (2015)
  doi:10.1088/1475-7516/2015/10/001
  [arXiv:1507.03875 [astro-ph.CO]].
  
\bibitem{ES}
J.~K.~Erickson, D.~H.~Wesley, P.~J.~Steinhardt and N.~Turok,
  ``Kasner and mixmaster behavior in universes with equation of state w >= 1,''
  Phys.\ Rev.\ D {\bf 69}, 063514 (2004)
  doi:10.1103/PhysRevD.69.063514
  [hep-th/0312009].
  
\bibitem{Peter}
Y.~F.~Cai, R.~Brandenberger and P.~Peter,
  ``Anisotropy in a Nonsingular Bounce,''
  Class.\ Quant.\ Grav.\  {\bf 30}, 075019 (2013)
  doi:10.1088/0264-9381/30/7/075019
  [arXiv:1301.4703 [gr-qc]].
  
\bibitem{LFI}
R.~H.~Brandenberger and J.~H.~Kung,
  ``Chaotic Inflation as an Attractor in Initial Condition Space,''
  Phys.\ Rev.\ D {\bf 42}, 1008 (1990).
  doi:10.1103/PhysRevD.42.1008;\\
H.~A.~Feldman and R.~H.~Brandenberger,
  ``Chaotic Inflation With Metric and Matter Perturbations,''
  Phys.\ Lett.\ B {\bf 227}, 359 (1989).
  doi:10.1016/0370-2693(89)90944-1;\\
W.~E.~East, M.~Kleban, A.~Linde and L.~Senatore,
  ``Beginning inflation in an inhomogeneous universe,''
  JCAP {\bf 1609}, 010 (2016)
  doi:10.1088/1475-7516/2016/09/010
  [arXiv:1511.05143 [hep-th]];\\
  K.~Clough, E.~A.~Lim, B.~S.~DiNunno, W.~Fischler, R.~Flauger and S.~Paban,
  ``Robustness of Inflation to Inhomogeneous Initial Conditions,''
  JCAP {\bf 1709}, 025 (2017)
  doi:10.1088/1475-7516/2017/09/025
  [arXiv:1608.04408 [hep-th]];\\
R.~Brandenberger,
  ``Initial conditions for inflation -- A short review,''
  Int.\ J.\ Mod.\ Phys.\ D {\bf 26}, no. 01, 1740002 (2016)
  doi:10.1142/S0218271817400028
  [arXiv:1601.01918 [hep-th]].
   
\bibitem{Piran}
D.~S.~Goldwirth and T.~Piran,
  ``Initial conditions for inflation,''
  Phys.\ Rept.\  {\bf 214}, 223 (1992).
  doi:10.1016/0370-1573(92)90073-9
  
\bibitem{PBB}
M.~Gasperini and G.~Veneziano,
  ``Pre - big bang in string cosmology,''
  Astropart.\ Phys.\  {\bf 1}, 317 (1993)
  doi:10.1016/0927-6505(93)90017-8
  [hep-th/9211021].

 \bibitem{Ekpflucts}
D.~H.~Lyth, 
``The Primordial curvature perturbation in the ekpyrotic universe,'' 
Phys.\ Lett.\ B {\bf 524}, 1 (2002)
doi:10.1016/S0370-2693(01)01374-0 [hep-ph/0106153];\\ 
  R.~Brandenberger and F.~Finelli, 
  ``On the spectrum of fluctuations in an effective field theory of the Ekpyrotic universe,'' 
  JHEP {\bf 0111}, 056 (2001) 
  doi:10.1088/1126-6708/2001/11/056 [hep-th/0109004].
  
\bibitem{KOST2}
J.~Khoury, B.~A.~Ovrut, P.~J.~Steinhardt and N.~Turok,
``Density perturbations in the ekpyrotic scenario,'' 
Phys.\ Rev.\ D {\bf 66}, 046005 (2002) 
doi:10.1103/PhysRevD.66.046005 [hep-th/0109050];\\
S.~Gratton, J.~Khoury, P.~J.~Steinhardt and N.~Turok,
  ``Conditions for generating scale-invariant density perturbations,''
  Phys.\ Rev.\ D {\bf 69}, 103505 (2004)
  doi:10.1103/PhysRevD.69.103505
  [astro-ph/0301395].
  
\bibitem{newEkp} 
A.~Notari and A.~Riotto, ``Isocurvature perturbations in the ekpyrotic universe,'' 
Nucl.\ Phys.\ B {\bf 644}, 371 (2002)
doi:10.1016/S0550-3213(02)00765-4 
[hep-th/0205019];\\ 
  F.~Finelli,
   ``Assisted contraction,'' 
   Phys.\ Lett.\ B {\bf 545}, 1 (2002) 
  doi:10.1016/S0370-2693(02)02554-6 
  [hep-th/0206112];\\
  F.~Di Marco, F.~Finelli and R.~Brandenberger,
  ``Adiabatic and isocurvature perturbations for multifield generalized Einstein models,''
  Phys.\ Rev.\ D {\bf 67}, 063512 (2003)
  doi:10.1103/PhysRevD.67.063512
  [astro-ph/0211276];\\
  J.~L.~Lehners, P.~McFadden, N.~Turok and P.~J.~Steinhardt,
  ``Generating ekpyrotic curvature perturbations before the big bang,''
  Phys.\ Rev.\ D {\bf 76}, 103501 (2007) 
  doi:10.1103/PhysRevD.76.103501
  [hep-th/0702153 [HEP-TH]];\\ 
E.~I.~Buchbinder, J.~Khoury and B.~A.~Ovrut, 
``New Ekpyrotic cosmology,'' Phys.\ Rev.\ D {\bf 76}, 123503 (2007)
doi:10.1103/PhysRevD.76.123503 [hep-th/0702154];\\ 
  P.~Creminelli and L.~Senatore, 
  ``A Smooth bouncing cosmology with scale invariant spectrum,'' 
  JCAP {\bf 0711}, 010 (2007)
  doi:10.1088/1475-7516/2007/11/010 
  [hep-th/0702165].;\\
A.~Ijjas, J.~L.~Lehners and P.~J.~Steinhardt,
  ``General mechanism for producing scale-invariant perturbations and small non-Gaussianity in ekpyrotic models,''
  Phys.\ Rev.\ D {\bf 89}, no. 12, 123520 (2014)
  doi:10.1103/PhysRevD.89.123520
  [arXiv:1404.1265 [astro-ph.CO]];\\
  A.~Fertig, J.~L.~Lehners, E.~Mallwitz and E.~Wilson-Ewing,
  ``Converting entropy to curvature perturbations after a cosmic bounce,''
  JCAP {\bf 1610}, 005 (2016)
  doi:10.1088/1475-7516/2016/10/005
  [arXiv:1607.05663 [hep-th]].

\bibitem{Khoury}
J.~Khoury and P.~J.~Steinhardt,
  ``Adiabatic Ekpyrosis: Scale-Invariant Curvature Perturbations from a Single Scalar Field in a Contracting Universe,''
  Phys.\ Rev.\ Lett.\  {\bf 104}, 091301 (2010)
  doi:10.1103/PhysRevLett.104.091301
  [arXiv:0910.2230 [hep-th]];\\
J.~Khoury and P.~J.~Steinhardt,
  ``Generating Scale-Invariant Perturbations from Rapidly-Evolving Equation of State,''
  Phys.\ Rev.\ D {\bf 83}, 123502 (2011)
  doi:10.1103/PhysRevD.83.123502
  [arXiv:1101.3548 [hep-th]].
  
\bibitem{DV}
R.~Durrer and F.~Vernizzi, 
``Adiabatic perturbations in pre - big bang models: Matching conditions and scale invariance,'' 
Phys.\ Rev.\ D {\bf 66}, 083503 (2002) 
doi:10.1103/PhysRevD.66.083503
[hep-ph/0203275];\\
C.~Cartier, R.~Durrer and E.~J.~Copeland,
  ``Cosmological perturbations and the transition from contraction to expansion,''
  Phys.\ Rev.\ D {\bf 67}, 103517 (2003)
  doi:10.1103/PhysRevD.67.103517
  [hep-th/0301198].

\bibitem{Tolley}
A.~J.~Tolley, N.~Turok and P.~J.~Steinhardt,
  ``Cosmological perturbations in a big crunch / big bang space-time,''
  Phys.\ Rev.\ D {\bf 69}, 106005 (2004)
  doi:10.1103/PhysRevD.69.106005
  [hep-th/0306109].
  
\bibitem{AdS}
R.~H.~Brandenberger, E.~G.~M.~Ferreira, I.~A.~Morrison, Y.~F.~Cai, S.~R.~Das and Y.~Wang,
  ``Fluctuations in a cosmology with a spacelike singularity and their gauge theory dual description,''
  Phys.\ Rev.\ D {\bf 94}, no. 8, 083508 (2016)
  doi:10.1103/PhysRevD.94.083508
  [arXiv:1601.00231 [hep-th]];\\
E.~G.~M.~Ferreira and R.~Brandenberger,
  ``Holographic Curvature Perturbations in a Cosmology with a Space-Like Singularity,''
  JCAP {\bf 1607}, 030 (2016)
  doi:10.1088/1475-7516/2016/07/030
  [arXiv:1602.08152 [hep-th]].
  
\bibitem{Galileon}
D.~A.~Easson, I.~Sawicki and A.~Vikman,
  ``G-Bounce,''
  JCAP {\bf 1111}, 021 (2011)
  doi:10.1088/1475-7516/2011/11/021
  [arXiv:1109.1047 [hep-th]];\\
A.~Ijjas and P.~J.~Steinhardt,
  ``Fully stable cosmological solutions with a non-singular classical bounce,''
  Phys.\ Lett.\ B {\bf 764}, 289 (2017)
  doi:10.1016/j.physletb.2016.11.047
  [arXiv:1609.01253 [gr-qc]];\\
  A.~Ijjas and P.~J.~Steinhardt,
  ``Classically stable nonsingular cosmological bounces,''
  Phys.\ Rev.\ Lett.\  {\bf 117}, no. 12, 121304 (2016)
  doi:10.1103/PhysRevLett.117.121304
  [arXiv:1606.08880 [gr-qc]];\\
  D.~A.~Dobre, A.~V.~Frolov, J.~T.~G.~Ghersi, S.~Ramazanov and A.~Vikman,
  ``Unbraiding the Bounce: Superluminality around the Corner,''
  JCAP {\bf 1803}, 020 (2018)
  doi:10.1088/1475-7516/2018/03/020
  [arXiv:1712.10272 [gr-qc]].
  
\bibitem{distance}
 H.~Ooguri and C.~Vafa,
 ``On the Geometry of the String Landscape and the Swampland,''
 Nucl.\ Phys.\ B {\bf 766}, 21 (2007)
 doi:10.1016/j.nuclphysb.2006.10.033
 [hep-th/0605264].
 
 \bibitem{Watson}
 S.~Watson,
  ``Moduli stabilization with the string Higgs effect,''
  Phys.\ Rev.\ D {\bf 70}, 066005 (2004)
  doi:10.1103/PhysRevD.70.066005
  [hep-th/0404177];\\
 L.~Kofman, A.~D.~Linde, X.~Liu, A.~Maloney, L.~McAllister and E.~Silverstein,
  ``Beauty is attractive: Moduli trapping at enhanced symmetry points,''
  JHEP {\bf 0405}, 030 (2004)
  doi:10.1088/1126-6708/2004/05/030
  [hep-th/0403001].
  
 \bibitem{Kounnas}
 R.~H.~Brandenberger, C.~Kounnas, H.~Partouche, S.~P.~Patil and N.~Toumbas,
  ``Cosmological Perturbations Across an S-brane,''
  JCAP {\bf 1403}, 015 (2014)
  doi:10.1088/1475-7516/2014/03/015
  [arXiv:1312.2524 [hep-th]];\\
  C.~Kounnas, H.~Partouche and N.~Toumbas,
  ``S-brane to thermal non-singular string cosmology,''
  Class.\ Quant.\ Grav.\  {\bf 29}, 095014 (2012)
  doi:10.1088/0264-9381/29/9/095014
  [arXiv:1111.5816 [hep-th]];\\
  C.~Kounnas, H.~Partouche and N.~Toumbas,
  ``Thermal duality and non-singular cosmology in d-dimensional superstrings,''
  Nucl.\ Phys.\ B {\bf 855}, 280 (2012)
  doi:10.1016/j.nuclphysb.2011.10.010
  [arXiv:1106.0946 [hep-th]].

\bibitem{MukhSas}
 M.~Sasaki,
 ``Large Scale Quantum Fluctuations in the Inflationary Universe,'' 
Prog.\ Theor.\ Phys.\  {\bf 76}, 1036 (1986).
doi:10.1143/PTP.76.1036;\\
V.~F.~Mukhanov, 
``Quantum Theory of Gauge Invariant Cosmological Perturbations,'' 
Sov.\ Phys.\ JETP {\bf 67}, 1297 (1988)
[Zh.\ Eksp.\ Teor.\ Fiz.\  {\bf 94N7}, 1 (1988)]. 
 
\bibitem{GSW}
M.~B.~Green, J.~H.~Schwarz and E.~Witten,
  ``Superstring Theory. Vol. 1: Introduction,''
  Cambridge, Uk: Univ. Pr. ( 1987) 469 P. ( Cambridge Monographs On Mathematical Physics);\\
  M.~B.~Green, J.~H.~Schwarz and E.~Witten,
  ``Superstring Theory. Vol. 2: Loop Amplitudes, Anomalies And Phenomenology,''
  Cambridge, Uk: Univ. Pr. ( 1987) 596 P. ( Cambridge Monographs On Mathematical Physics).
   
\bibitem{Patil}
R.~H.~Brandenberger, A.~Nayeri, S.~P.~Patil and C.~Vafa,
  ``String gas cosmology and structure formation,''
  Int.\ J.\ Mod.\ Phys.\ A {\bf 22}, 3621 (2007)
  doi:10.1142/S0217751X07037159
  [hep-th/0608121];\\
  R.~H.~Brandenberger, A.~Nayeri and S.~P.~Patil,
  ``Closed String Thermodynamics and a Blue Tensor Spectrum,''
  Phys.\ Rev.\ D {\bf 90}, no. 6, 067301 (2014)
  doi:10.1103/PhysRevD.90.067301
  [arXiv:1403.4927 [astro-ph.CO]].
                       
\end{thebibliography}
\end{document}